\documentclass[showpacs,aps,prd,nofootinbib,floatfix,amsmath,amssymb]{revtex4}
\usepackage{graphicx}
\makeatletter

\begin{document}

\title{CP violating effects in the decay $Z \to \mu^+\mu^-\gamma$
induced by $ZZ\gamma$ and $Z\gamma\gamma$ couplings}
\author{M. A. P\' erez}
\email[E-mail:]{mperez@fis.cinvestav.mx}
\author{F. Ram\'\i rez-Zavaleta}
\email[E-mail:]{rzf@fis.cinvestav.mx} \affiliation{Departamento de
F\'\i sica, CINVESTAV, Apartado Postal 14-740, 07000, M\'exico, D.
F., M\'exico}

\date{\today}

\begin{abstract}
We analyze possible CP-violating effects induced in the $Z$ decay
with hard photon radiation by $\gamma ZZ$ and $\gamma\gamma Z$
anomalous vertices. We estimate the sensibility of future linear
collider experiments on these couplings coming from CP-odd
asymmetries associated to angular correlations of the three particle
final state in $e^+e^- \to Z \to \mu^+\mu^-\gamma$. We find that a
linear collider with an integrated luminosity of 500 $fb^{-1}$ and
$\sqrt{s} = 0.05$ TeV can place the bound $|h_1^{\gamma,Z}| < 0.92$
at the $90\%$ confidence level for these couplings.
\end{abstract}


\maketitle

\label{int}Trilinear gauge boson self-couplings have been studied as
a new source of CP-violating interactions at high energies
\cite{QUE, HAG, LAR, XIO}. In particular, anomalous neutral triple
gauge couplings (NTGC), which are not present at tree level in the
Standard Model (SM), may induce CP-violating effects in processes
such as $\gamma e \to Ze$ \cite{CHO}, $e^+e^- \to \gamma Z$
\cite{HUR}, $e^+e^- \to ZZ$ \cite{CHA}, $\gamma\gamma \to t\bar{t}$
\cite{POU} and $p\bar{p} \to Z\gamma$ \cite{XIO}.

In the present paper we are interested in studying the CP-violating
effects induced by NTGC in the radiative decay $Z \to
\mu^+\mu^-\gamma$. The contributing diagrams to this decay mode are
depicted in Fig. $1$. Fig. $1(a)$ refers to the SM contribution, the
other diagrams include CP-violating effective vertices denoted by
heavy dots. To lowest order, the CP-violating effects in this
reaction are induced by the interference between the SM contribution
and any one of the other diagrams shown in Fig. 1. Figs. 1(b) and
1(c) correspond to the contributions arising from the electric and
weak dipole moments of the muon. Since these moments are tightly
bounded \cite{EXP}, their contribution exclude any observable
CP-violating effect in the decay $Z \to \mu^+\mu^-\gamma$ \cite{XIO,
BER}. The possible CP-violating effects induced by the contact
interaction $\mu\mu\gamma Z$ shown in fig. 1(d) have been studied in
Ref. \cite{BER}. However, it has been pointed out \cite{XIO, TRA}
that a fully gauge invariant effective Lagrangian that induces this
local interaction also generates the CP-violating vertices $\mu\mu
Z$ and $\mu\mu\gamma$ shown in Fig. 1(b) and 1(c). When all these
new vertices are taken into account, the interference between the
lowest order SM diagram and the new CP-violating contributions
vanishes for the process $e^+e^- \to \mu^+\mu^-\gamma$ and one is
left with no CP-odd effects. Nonetheless, the L$3$ collaboration
searched for CP-violating effects in the decay of the $Z$ boson with
hard photon radiation and they could set an upper bound on the
CP-violating coupling $\mu^+\mu^-\gamma Z$ \cite{aci}.

We will assume that possible CP-violating effects in the decay $Z
\to \mu^+\mu^-\gamma$ are induced by the effective vertices
$ZZ\gamma$ and $Z\gamma\gamma$ shown in Fig. 1(e). In Ref.
\cite{BER}, it was argued that is not necessary to study the
CP-violating effects induced in this decay mode by the NTGC since
the Lorentz structure of the respective amplitude reduces to that
obtained for the contact interaction $\mu^+\mu^-\gamma Z$, Fig.
1(d). Our point of view in the present paper is that this argument
is not enough to disregard the contribution of the TNGC to the
CP-violating effects in the decay $Z \to \mu^+\mu^-\gamma$. It has
been pointed out \cite{LAR, ALC} that these effective vertices have
a rather rich structure in the framework of the effective Lagrangian
formalism. They may be generated in both the linear and non-linear
realization of the $SU(2)_L \times U(1)$ gauge symmetry. We have
shown \cite{LAR} that in both realizations, the NTGC induce the same
Lorentz structure but that this structure may be induced by
operators of dimension $6$ in the non-linear scenario or operators
of dimension $8$ in the case of the linear realization of the
$SU(2)_L \times U(1)$ symmetry. As a consequence, any one of these
constructions may generate CP-odd observables which could be
independent of those induced by the $\mu^+\mu^-\gamma Z$ contact
interaction or the weak and electric dipole moments of the muon.

\begin{figure}
\centering
\includegraphics[width=5in]{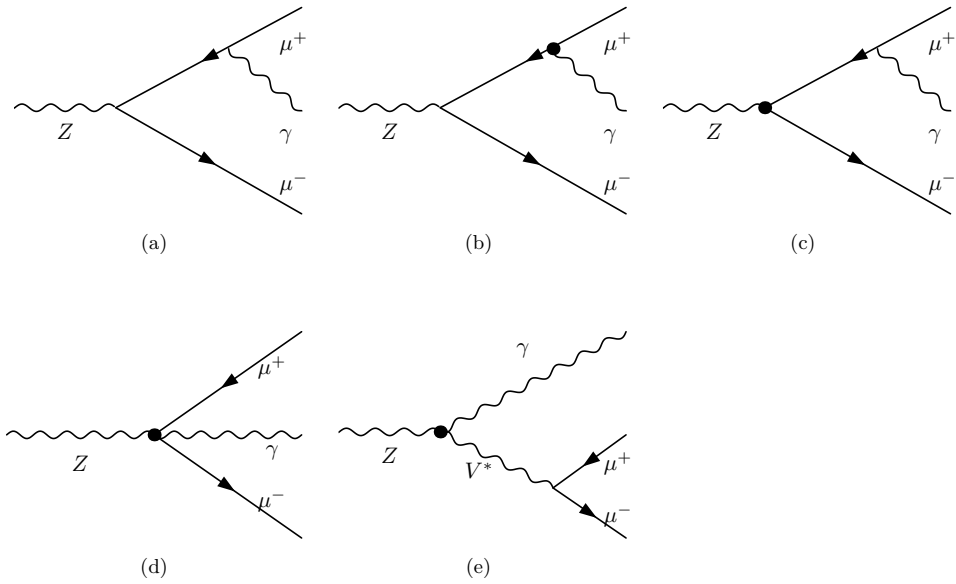}
\caption{\label{FIG0}.Feynman diagrams contributing to CP-violating
effects in the decay $Z \to \mu^+\mu^-\gamma$. The heavy dots denote
CP-violating effective vertices. Crossed diagrams are not shown,
$V^*$ stands for $Z$ or $\gamma$.}
\end{figure}

\begin{figure}
\centering
\includegraphics[width=3.9in]{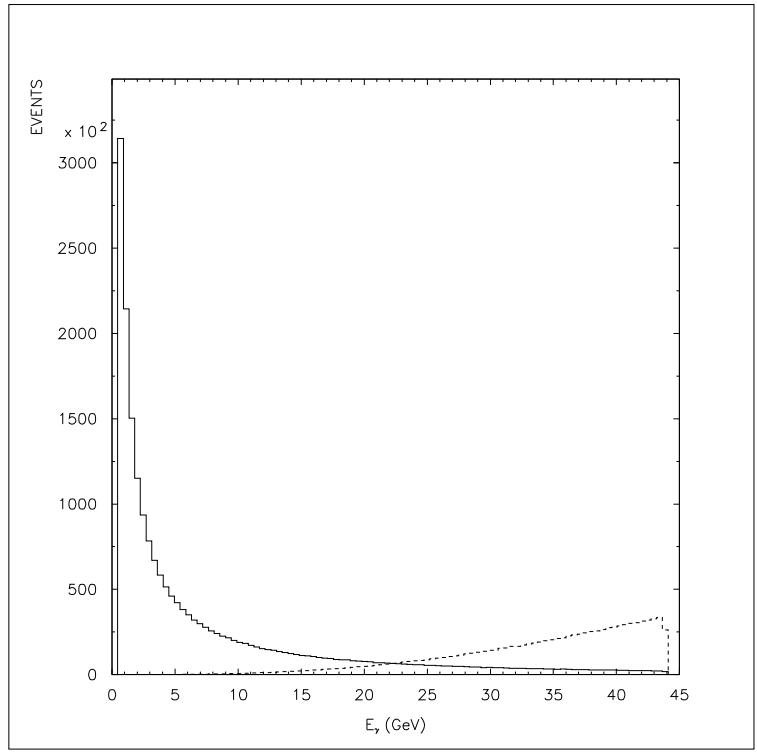}
\caption{\label{FIG1}Energy distributions of the photon emitted in
$e^+e^- \to \mu^+\mu^-Z$ for SM (solid line) and TNGB (dotted line)
contributions with $h_1^V=1$.}
\end{figure}

The most general form of the $Z^\alpha(p)V^\beta(q)\gamma^\mu(k)$
vertex function, with $V^\beta = \gamma, Z$, when $Z^\alpha$ and
$\gamma^\mu$ are on-shell and which respects Lorentz and
electromagnetic gauge invariance is given by \cite{HAG, LAR}
\begin{eqnarray}\label{frvtbnn}
\Gamma_{\alpha\beta\mu}^{ZV^{*}\gamma}(p,q,k)&=&\frac{ie}{m_{Z}^{2}}
[h_{1}^{V}(k^{\alpha}g^{\mu\beta}-k^{\beta}g^{\mu\alpha})+
\frac{h_{2}^{V}}{m_{Z}^{2}}q^{\alpha}(q \cdot k
g^{\beta\mu}-k^{\beta}q^{\mu})\nonumber\\& &+
h_{3}^{V}\epsilon^{\alpha\beta\mu\nu}k_{\nu}+
\frac{h_{4}^{V}}{m_{Z}^{2}}q^{\alpha}\epsilon^{\beta\mu\nu\rho}p_{\nu}q_{\rho}]
(q^{2}_V-m_{V}^{2}),
\end{eqnarray}
where $m_Z$ is the $Z$-boson mass and the first two terms in
(\ref{frvtbnn}) are CP-violating and the other two are
CP-conserving. The respective couplings $h_i^{\gamma,Z}$ have been
bounded in $e^+e^-$ and $p\bar{p}$ collisions \cite{EXP, WYN}:
$h_1^Z \in [-0_\cdot15, 0_\cdot14]$, $h_2^Z \in [-0_\cdot09,
0_\cdot08]$, $h_3^Z \in [-0_\cdot22, 0_\cdot11]$, $h_4^Z \in
[-0_\cdot07, 0_\cdot15]$, $h_1^\gamma \in [-0_\cdot06, 0_\cdot06]$,
$h_2^\gamma \in [-0_\cdot05, 0_\cdot02]$, $h_3^\gamma \in
[-0_\cdot06, 0_\cdot004]$, $h_4^\gamma \in [-0_\cdot004,
0_\cdot042]$.

\begin{figure}
\centering
\includegraphics[width=3.9in]{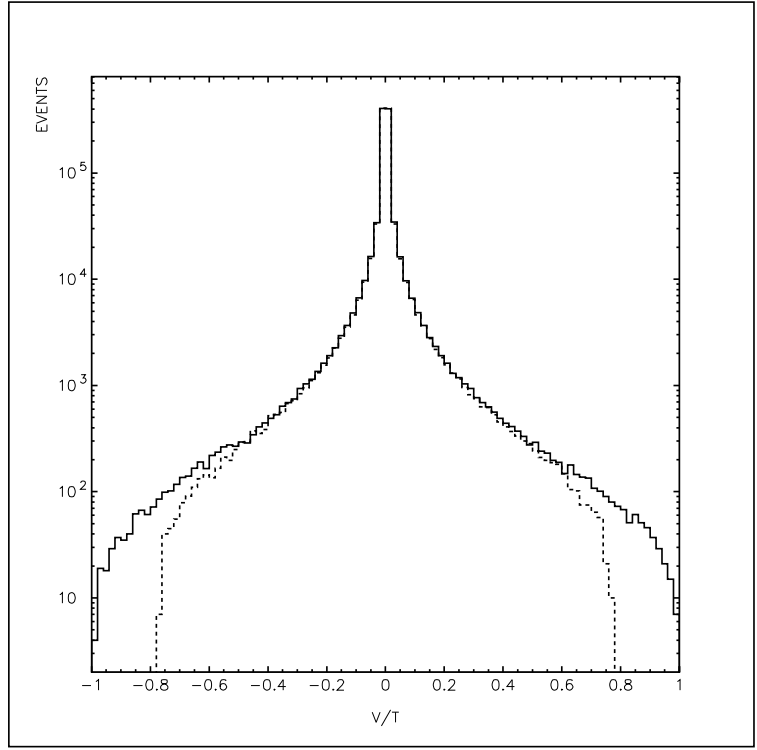}
\caption{\label{FIG3}Distributions for the CP-odd observables $V$
(solid line) and $T$ (dotted line) corresponding to the interference
between the SM and TNGB contributions ($h_1^V=1$).}
\end{figure}

\begin{figure}
\centering
\includegraphics[width=3.9in]{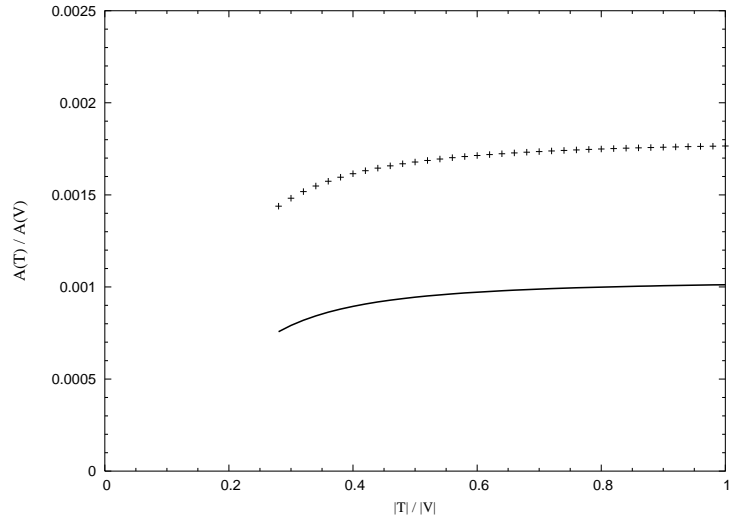}
\caption{\label{FIG6}Asymmetries for the CP-odd observables $V$
(dotted line) and $T$ (solid line) corresponding to the interference
between the SM and TNGB contributions ($h_1^V=1$).}
\end{figure}

\begin{figure}
\centering
\includegraphics[width=3.9in]{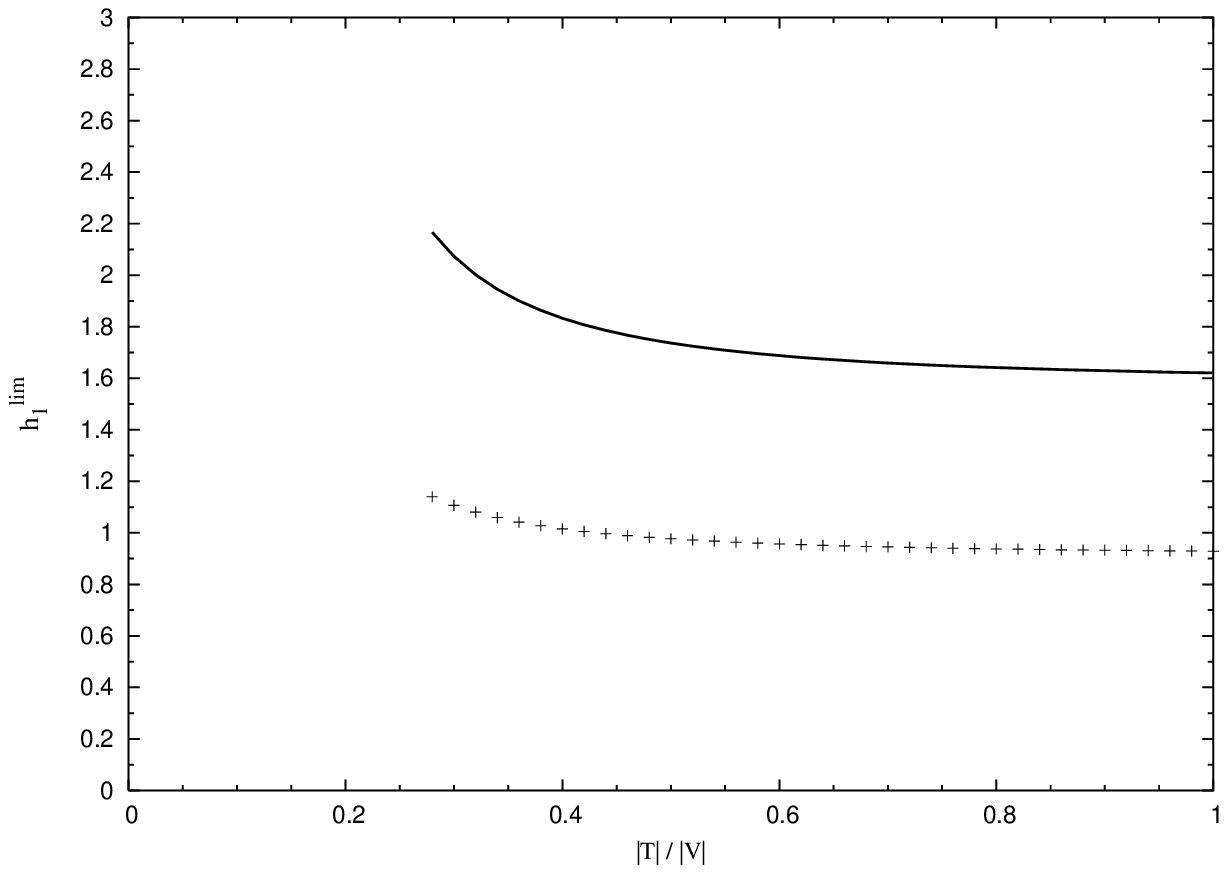}
\caption{\label{FIG8}The $90$\% C. L. contours on $h_1^V$ from the
asymmetries $A_T$ (solid line) and $A_V$ (dotted line) as function
of the CP-odd variables $T$ and $V$.}
\end{figure}

Following Ref. \cite{BER, aci}, we will use the angular correlations
of the three particles in the final state in $e^+e^- \to Z \to
\mu^+\mu^-\gamma$ in order to test CP-violating effects induced by
the anomalous $ZZ\gamma/Z\gamma\gamma$ couplings. The following
CP-odd observables have been proposed \cite{BER} to search for
CP-violation in this radiative decay
\begin{eqnarray}\label{TVvariables}
T&=&(\widehat{\textbf{k}}_{\mu^{+}}-\widehat{\textbf{k}}_{\mu^{-}})
\cdot\widehat{\textbf{p}}_{e^{+}}(\widehat{\textbf{k}}_{\mu^{+}}
\times\widehat{\textbf{k}}_{\mu^{-}})\cdot\widehat{\textbf{p}}_{e^{+}}\nonumber\\
V&=&(\widehat{\textbf{k}}_{\mu^{+}}
\times\widehat{\textbf{k}}_{\mu^{-}})\cdot\widehat{\textbf{p}}_{e^{+}}.
\end{eqnarray}
where $\widehat{\textbf{k}}_{\mu^{+}}$ and
$\widehat{\textbf{k}}_{\mu^{-}}$ are both the unitary vectors of
muon and antimuon, respectively, $\widehat{\textbf{p}}_{e^{+}}$ is
the direction of colliding positron.

We use the narrow-width approximation and in order to suppress the
SM contributions from radiative $Z$ decay, we impose the cuts: $P_{T
\gamma} \geq 10$ GeV, $P_{T\mu^-} \geq 20$ GeV, $P_{T\mu^+} \geq 20$
GeV \cite{XIO, aci}. In Fig. \ref{FIG1} we present the energy
distributions for the emitted photon when we consider separately the
SM and TNGB contributions. While the SM contribution is peaked in
the forward direction, the CP-odd distribution has a maximum at
large values of the photon energy. Fig. \ref{FIG3} shows the
distributions expected in this $Z$ radiative decay for the
observables $T$ and $V$ obtained with a Monte Carlo program for the
numerical integration using $h_1^V$. It is observed a rather slight
asymmetry between negative and positive values of $T$ and $V$. In
order to estimate the sensitivity of the TESLA collider to the
$h_1^V$ couplings, we will make use of the relative asymmetries
between negative and positive values of the observables $T$ and $V$
\cite{aci}:
\begin{eqnarray}\label{asy1}
A_T &=& \frac{N_{T>0}-N_{T<0}}{N_{T>0}+N_{T<0}}\\
\label{asy2} A_V &=& \frac{N_{V>0}-N_{V<0}}{N_{V>0}+N_{V<0}}.
\end{eqnarray}

In Fig. \ref{FIG6} these asymmetries are shown for intervals in
$|T|$ and $|V|$, respectively. Using these results, we can obtain
$90$ \% CL limits on the $h^V_1$ vertices as a function of $|T|$ and
$|V|$ (Fig. \ref{FIG8}). We find that the best limits are obtained
for $|V|\sim 1$, which correspond to $|h_1^V| \leq 0.92$. These
limits correspond to a future linear collider with $\sqrt{s}=500$
GeV and an integrated luminosity of 500 $fb^{-1}$. We found no
difference for the asymmetries induced by the $h_1^Z$ and
$h_1^\gamma$ couplings due to the Lorentz structure shown in Eq.
(\ref{frvtbnn}), while the respective CP-odd effects induced by the
couplings $h_2^V$ are even more suppressed due to the extra
suppression factor $m_Z^2$ in Eq. (\ref{frvtbnn}).

In conclusion, our numerical study has shown that anomalous
CP-violating vertices $ZZ\gamma$ and $Z\gamma\gamma$ also give rise
to CP-odd asymmetries associated to angular correlations of the
three particle final state. While rare $Z$ decays may be sensitive
to both the CP-conserving and CP-violating $h_i^V$ couplings
\cite{PER}, the detection of the $A_T$ and $A_V$ asymmetries in the
radiative decay $Z \to \mu^+\mu~-\gamma$ may be used to get specific
constraints on the CP-violating $h_i^V$ couplings. We have derived
sensitivity limits expected in a future linear collider for the
CP-violating vertices $ZZ\gamma$ and $Z\gamma\gamma$. These
sensitivity limits on $h_1^V$ are about one order of magnitude below
than those obtained from CP-conserving observables in $e^+e^-$
experiments \cite{WYN, ATA} or from future experiments in $e^+e^-
\to \gamma Z$ with a transverse beam polarization \cite{CHA}.
However, the analysis presented in this paper has shown that an
improvement in the luminosity expected in a future TeV linear
collider \cite{ELL} may impose better and independent bounds on the
CP-odd couplings $h_1^V$.

The calculations involved in the present work needed a considerable
symbolic and numerical computation. We achieved this objective by
using FeynCalc into the Mathematica for symbolic manipulations as
reduction of traces of Dirac matrices, simplification of large
expressions, etc. Fortran program code was used as translator from
results obtained by VEGAS to graphical environment by program PAW++.
The numerical phase space integrations were done with VEGAS program
into the COMPHEP, like also Monte Carlo events generation.

\acknowledgments{We acknowledge support from Conacyt and SNI (M\'
exico). We would like to thank G. Tavares-Velasco for collaboration
in the early stages of the present work and H. Castilla and A.
S\'anchez for useful discussions.}

\end{document}